\newif\ifsections
\sectionsfalse
\newif\ifpnas
\pnastrue

\ifpnas
  \documentclass[9pt,twoside]{pnas-new-nolines}
  \templatetype{pnasmathematics}
\else
  \documentclass[12pt]{article}
\fi  

\usepackage{scrtime} 
\date{\today\ @ \thistime}

\usepackage{pdflscape}

\usepackage{MnSymbol} 

\usepackage{placeins} 

\ifpnas
 \usepackage{lineno}
\else 
 \usepackage{lineno}
\fi

\DeclareMathAlphabet{\mathdutchcal}{U}{dutchcal}{m}{n} 

\usepackage{amsmath,amsfonts,enumerate,float,fullpage,graphicx,longtable,lscape,mathtools,multirow,rotating,stmaryrd,subcaption,xcolor}

\usepackage{xspace}

\usepackage{tabularx}
\usepackage{dcolumn}
\newcolumntype{d}[1]{D{.}{.}{#1}}

\usepackage{hyperref}
\usepackage[nameinlink,capitalize]{cleveref}
\crefformat{section}{\S#2#1#3}
\crefname{section}{\S}{\S\S}
\Crefname{section}{Section}{Sections}
\ifpnas
  \crefname{equation}{Eq.}{Eqs.}
  \Crefname{equation}{Eq.}{Eqs.}
  \crefname{figure}{Fig.}{Figs.}
  \Crefname{figure}{Fig.}{Figs.}
  \renewcommand{\eqref}[1]{\textup{{\normalfont[\ref{#1}}\normalfont]}}
  \crefformat{equation}{Eq.\,#2\textbf{#1}#3}
  \Crefformat{equation}{Eq.\,#2\textbf{#1}#3}
  \crefmultiformat{equation}
        {Eq.~#2\textbf{#1}#3}
        { and~#2\textbf{#1}#3}
        {, #2\textbf{#1}#3}
        { and~#2\textbf{#1}#3}
  \crefrangeformat{equation}{Eqs.~#3\textbf{#1}#4--#5\textbf{#2}#6}
  \Crefrangeformat{equation}{Eqs.~#3\textbf{#1}#4--#5\textbf{#2}#6}
\else
  \crefname{equation}{Equation}{Equations}
  \Crefname{equation}{Equation}{Equations}
  \crefname{figure}{Figure}{Figures}
  \Crefname{figure}{Figure}{Figures}
\fi
\Crefname{appsec}{Appendix}{appendices}
\let\eqref=\labelcref

\usepackage[all,color]{xy}
\usepackage{xcolor}
\definecolor{Scol}{HTML}{4DAF4A}
\definecolor{Icol}{HTML}{E41A1C}
\definecolor{Rcol}{HTML}{377EB8}

\usepackage{framed} 
\definecolor{shadecolor}{rgb}{0.9,0.9,0.9}

\usepackage{xcolor}
\definecolor{oiorange}{RGB}{230,105,0}        
\definecolor{oiskyblue}{RGB}{86,180,233}      
\definecolor{oibluishgreen}{RGB}{0,158,115}   
\definecolor{oiyellow}{RGB}{240,228,66}       
\definecolor{oiblue}{RGB}{0,114,178}          
\definecolor{oivermillion}{RGB}{213,94,0}     
\definecolor{oireddishpurple}{RGB}{204,121,167} 
\definecolor{oigray}{RGB}{153,153,153}        

\newcommand{\ie}{\textit{i.e., }}
\newcommand{\eg}{\textit{e.g., }}

\newcommand{\dee}{{\rm d}}
\newcommand{\dd}[2]{{\frac{\dee{#1}}{\dee{#2}}}}

\newcommand{\ddt}[1]{\dd{#1}{t}}

\newcommand{\tinit}{t_{\rm i}}

\newcommand{\xvecinit}{{\vec{x}}_{\rm i}}
\newcommand{\Sinit}{S_{\rm i}}

\newcommand{\raiseboxamount}{0.35ex}
\newcommand{\DFEnoraise}{\scalebox{0.5}{\textnormal{\textsc{dfe}}}}
\newcommand{\DFE}{\raisebox{\raiseboxamount}{\DFEnoraise}}
\newcommand{\SDFE}{\displaystyle S^{\DFE}}
\newcommand{\xvecDFE}{\xxx^{\DFE}}
\newcommand{\EEnoraise}{\scalebox{0.5}{\textnormal{\textsc{ee}}}}
\newcommand{\EE}{\raisebox{\raiseboxamount}{\EEnoraise}}
\newcommand{\SEE}{\displaystyle S^{\EE}}
\newcommand{\xEE}{x^{\EEnoraise}}
\newcommand{\xEEi}{\xEE_i}
\newcommand{\xEEj}{\xEE_j}
\newcommand{\xEEk}{\xEE_k}
\newcommand{\xEEl}{\xEE_\ell}
\newcommand{\xvecEE}{\xxx^{\EE}}

\newcommand{\LiapEE}{\displaystyle L^{\!\EE}}
\newcommand{\LiapDFE}{\displaystyle L^{\!\DFE}}
\newcommand{\FoIEE}{\displaystyle\FoI^{\!\!\EE}}
\newcommand{\doublevec}[1]{\overset{\smash{\raisebox{-0.35ex}{$\scriptscriptstyle \twoheadrightarrow$}}}{#1}}
\newcommand{\Xvec}{\doublevec{X}}
\newcommand{\XDFE}{\Xvec^{\raisebox{-0.2ex}{\DFEnoraise}}}
\newcommand{\XEE}{\Xvec^{\raisebox{-0.2ex}{\EEnoraise}}}

\newcommand{\Aset}{{\mathcal A}}
\newcommand{\Bset}{{\mathcal B}}
\newcommand{\Dset}{{\mathcal D}}

\newcommand{\inc}{i}

\newcommand{\R}{{\mathcal R}}

\newcommand{\Rzero}{\Rnaught}

\usepackage{stackengine}

\newcommand{\FoI}{\Lambda}

\makeatletter
\newcommand{\vast}{\bBigg@{4}}
\newcommand{\Vast}{\bBigg@{5}}
\makeatother

\makeatletter
\def\semibig#1{{\hbox{$\left#1\vbox to11\p@{}\right.\n@space$}}}
\makeatother

\usepackage{mathrsfs} 

\newcommand{\nocomment}[3]{}

\newcommand{\term}[1]{{\emph{#1}}}

\newcommand{\Methods}{\hyperlink{Methods}{Methods}\xspace}

\usepackage[normalem]{ulem}
\newcommand{\stkout}[1]{{\color{blue}\ifmmode\text{\sout{\ensuremath{#1}}}\else\sout{#1}\fi}}

\newcommand{\supp}{\hyperlink{supp}{Supplementary Information}\xspace}

\newcommand{\mulatent}{\mu_{{}_{\rm L}}}
\newcommand{\mulatentsmallL}{\mu_{\text{\scalebox{0.7}{${}_{\rm L}$}}}}

\usepackage{amsthm}

\input xy
\xyoption{all}

\newcounter{Remark}                             
\newcounter{Definition}

\newtheorem{theorem}{Theorem}

\newtheorem*{lemma}{Lemma}

\numberwithin{equation}{section}

\newcounter{RemarkCtr}
\newcounter{ExCtr}

\newcommand{\eref}[1]{(\ref{#1})}

\newcounter{assCtr}

\def \Rzero{\mathcal R_0}
\def \R{\mathbb R}

\def \Rgez{\R_{\ge 0}}

\def \Rgtz{\R_{> 0}}

\def \zero{\vec{\bf 0}}
\def \zero{\vec{0}}
\def \ones{\vec{1}}
\def \eee{\textrm{e}}

\def \sgn{\text{\upshape sign}}

\def \aand{\quad \textrm{ \rm and } \quad}
\def \aaand{\hspace{1.5cm} \textrm{\rm and } \hspace{1.5cm}}

\def \xxx{\vec x}
\def \dxxxdt{\frac{\dee{} \xxx}{\dee{}t}}
\def \dSdt{\frac{\dee{}S}{\dee{}t}}

\def \dUdt{\displaystyle\frac{\dee{}\displaystyle\LiapDFE}{\dee{}t}}

\def \dVdt{\displaystyle\frac{\dee{}\displaystyle\LiapEE}{\dee{}t}}
\def \dNdt{\frac{\dee{}N}{\dee{}t}}

\def \PPP{\vec P}
\def \bbbeta{\vec \beta}

\def \FoI{\Lambda}		

\def \LL{\ell}
\def \ill{i'}

\def \sumjbeta{\sum_{j=1}^n \beta_j \xEEj \SEE}

\def \summkj{\sum_{j=1}^n m_{kj} \xEEj}

\def \={\;\; = \;\;}
\def \G{g}

\def \rr{\LL}

\def \AA{A}

\newcommand{\GG}[1]{\G \hspace{-0.06cm} \left( {#1} \right)}
\def \lambdaNaught{\lambda_0}
\def \rsrs{r}

\usepackage{placeins} 

\usepackage{tikz}
\usepackage{pgfplots}

\usepackage{hyperref}
\usepackage[nameinlink,capitalize]{cleveref}
\crefname{equation}{Equation}{Equations}
\Crefname{equation}{Equation}{Equations}
\crefname{figure}{Figure}{Figures}
\Crefname{figure}{Figure}{Figures}
\Crefname{appsec}{Appendix}{appendices}
\crefname{lemma}{lemma}{lemmas}
\Crefname{lemma}{Lemma}{Lemmas}
\crefname{assumption}{assumption}{assumptions}
\usepackage{enumerate}
\usepackage{xcolor}
\newcommand{\transpose}{{\textsf{\upshape T}}}
\newcommand{\susrec}{\nu}

\newcommand{\dtau}{\,\dee{\tau}}

\definecolor{Scol}{HTML}{4DAF4A}
\definecolor{Ecol}{HTML}{FFD92F}
\definecolor{Icol}{HTML}{E41A1C}
\definecolor{Rcol}{HTML}{377EB8}

\ifpnas
  \crefname{equation}{Eq.}{Eqs.}
  \Crefname{equation}{Eq.}{Eqs.}
  \crefname{figure}{Fig.}{Figs.}
  \Crefname{figure}{Fig.}{Figs.}
  \renewcommand{\eqref}[1]{(\cref{#1})}
\else
  \crefname{equation}{Equation}{Equations}
  \Crefname{equation}{Equation}{Equations}
  \crefname{figure}{Figure}{Figures}
  \Crefname{figure}{Figure}{Figures}
\fi
\Crefname{appsec}{Appendix}{appendices}

\def\G{g}

\renewcommand{\theequation}{\arabic{equation}}

\title{Global stability of epidemic models with uniform susceptibility}

\ifpnas
  \author[a,b,1]{David J.\,D.\ Earn}
  \author[c]{C.\ Connell McCluskey}

  \affil[a]{Department of Mathematics and Statistics, McMaster University, Hamilton, ON L8S 4K1, Canada}
  \affil[b]{M.G.\ DeGroote Institute for Infectious Disease Research, McMaster University, Hamilton, ON L8S 4L8, Canada}
  \affil[c]{Mathematics, Wilfrid Laurier University, Waterloo, ON N2L 3C5, Canada}

  \leadauthor{Earn}

\else
  \author{David J.\,D.\ Earn$^{1,2}$ and C.\ Connell McCluskey$^3$\\[6pt]
  \parbox{\textwidth}{\centering \small
    $^1$Department of Mathematics \& Statistics
    and $^2$M.\,G.\,DeGroote Institute for \\Infectious Disease Research,
    McMaster University, Hamilton, Ontario, Canada;\\
    {\footnotesize ORCID: 0000-0002-7562-1341, e-mail: earn@math.mcmaster.ca}\\[3pt]
    $^3$Department of Mathematics, Wilfrid Laurier University,
    Waterloo, Ontario, Canada;\\
    {\footnotesize ORCID: 0009-0005-9074-5106, e-mail: cmccluskey@wlu.ca}
    }
  }
\fi 

\ifpnas
  \significancestatement{
Mathematical models are widely used to study the spread of infectious diseases. A central question is whether these models predict that disease incidence will: oscillate indefinitely, settle down to a constant level, or behave unpredictably over time.  While dynamics near a steady state can often be determined straightforwardly, it is usually more challenging to establish behaviour from arbitrary starting conditions.  We prove---for a broad class of models, including thousands studied previously---that there is a \emph{unique}, stable equilibrium prevalence that is approached whenever the disease is present.  This result eliminates the need for further case-by-case analysis, and clarifies that such models cannot exhibit more complicated long-term behaviour, such as sustained oscillations or multiple stable outcomes.
  }

  \authorcontributions{D.J.D.E.\ and C.C.M.\ contributed equally to this paper, and each had input on all parts of the work.}
  \authordeclaration{The authors declare no competing interests.}
  \correspondingauthor{\textsuperscript{1}To whom correspondence should be addressed. \mbox{E-mail: earn@math.mcmaster.ca}}


\fi 

\let\citen = \cite

\makeatletter
\newcommand{\myPNASRightFooter}{%
  \footerfont
  {\bfseries PNAS} \quad 2025 \quad Vol.\ 122 \quad No.\ 49 \quad e2510156122 \quad
  \textbf{\thepage\ of \pageref{LastPage}}%
}
\fancypagestyle{firststyle}{%
  \fancyhead{}
  \fancyfoot[L]{\footerfont\@ifundefined{@doi}{}{\@doi}}%
  \fancyfoot[C]{}%
  \fancyfoot[R]{\myPNASRightFooter}%
}
\makeatother
\begin{document}

\setboolean{displaywatermark}{false}  

\ifpnas
  \relax
\else
  \maketitle
  \linenumbers
\fi

\begin{abstract}

Transmission dynamics of infectious diseases are often studied using compartmental mathematical models, which are commonly represented as systems of autonomous ordinary differential equations.  A key step in the analysis of such models is to identify equilibria and find conditions for their stability.  Local stability analysis reduces to a problem in linear algebra, but there is no general algorithm for establishing global stability properties.  Substantial progress on global stability of epidemic models has been made in the last 20 years, primarily by successfully applying Lyapunov's method to specific systems.  Here, we show that any compartmental epidemic model in which susceptible individuals cannot be distinguished and can be infected only once, has a globally asymptotically stable (GAS) equilibrium.  If the basic reproduction number $\Rzero$ satisfies $\Rzero>1$, then the GAS fixed point is an endemic equilibrium (\ie constant, positive disease prevalence).  Alternatively, if $\Rzero\le1$, then the GAS equilibrium is disease-free.  This theorem subsumes a large number of results published over the last century, strengthens most of them by establishing global rather than local stability, avoids the need for any stability analyses of these systems in the future, and settles the question of whether co-existing stable solutions or non-equilibrium attractors are possible in such models: they are not.

\end{abstract}

\ifpnas
  \dates{\color{black}This author-typeset version differs only in formatting from
  the paper as \href{https://www.pnas.org/cgi/doi/10.1073/pnas.2510156122}{published in PNAS}
  on December 2, 2025.
  The paper should be cited as\hfill\break
  DJD Earn, CC McCluskey,
  {Global stability of epidemic models with uniform susceptibility},
  \emph{Proc.\ Natl.\ Acad.\ Sci.\ U.S.A.} \textbf{122}(49), e2510156122 (2025).}
  \doi{\url{https://www.pnas.org/cgi/doi/10.1073/pnas.2510156122}}
  \maketitle
  \thispagestyle{firststyle}
  \ifthenelse{\boolean{shortarticle}}{\ifthenelse{\boolean{singlecolumn}}{\abscontentformatted}{\abscontent}}{}
\fi 

\ifpnas
  \firstpage[3]{3}

\fi

\ifpnas\relax\else\medbreak\fi
\ifsections
  \section{Introduction}\label{sec:intro}
\fi

\typeout{(1) Current \columnsep = \the\columnsep} 

\ifpnas
\fi

\ifpnas\dropcap{M}\else{M}\fi{}athematical modelling of infectious disease transmission dates
back to Daniel Bernoulli in the 18\textsuperscript{th} century
\cite{Bern1760}.  Compartmental epidemic models were first published
in the early 20\textsuperscript{th} century \cite{Hame06,Ross11},
which led to the landmark paper of Kermack and
McKendrick \cite{KermMcKe27} (KM) in 1927.  In its simplest form, the
KM model reduces to a system of three ordinary differential
equations (ODEs),
\begin{linenomath*}
\begin{align}\label{eq:KM}
  \ddt{S} \;=\; -\beta SI \,, \qquad
  \ddt{I} \;=\; \beta SI  - \gamma I \,, \qquad
  \ddt{R} \;=\; \gamma I \,,
\end{align}
\end{linenomath*}
where $S$, $I$, and $R$ are the numbers of susceptible, infected, and
recovered hosts, and the total population size is $N=S+I+R$.  In
this \term{SIR model}, the parameters, $\beta$ and $\gamma$, are the
rates of transmission and recovery; the mean infectious period is
$1/\gamma$ and the basic reproduction number is
$\Rzero=N\beta/\gamma$.  The \term{incidence function},
$\inc(S,I)=\beta SI$, is linear in $S$ and $I$, representing
homogeneous mixing of susceptible and infected individuals, as if they
were elements of an ideal gas.  There is no flow from $I$ or $R$ back
to $S$, so infected individuals become permanently immune upon
recovery.

Because there is no source of new susceptibles in \cref{eq:KM}, any
initial state yields at most one outbreak, and ultimately converges on
a \term{disease-free equilibrium} (DFE) at which $I=0$.
However, \cref{eq:KM} ignores \term{vital dynamics} (births and
deaths).  In the more realistic situation that includes vital
dynamics, $\Rzero$ depends on the vital rates and, if $\Rzero>1$,
there is an \term{endemic equilibrium} (EE) at which $I>0$.  It has
been known for at least 50 years that the SIR model with vital
dynamics always has a \term{globally asymptotically stable} (GAS)
equilibrium: if $\Rzero\le1$ then all solutions converge to the DFE,
and if $\Rzero>1$ then all initial states (except those with $I=0$)
yield solutions that converge to the EE
\cite{Heth76}.

\begin{figure}
\begin{center}
\includegraphics[width=\textwidth]{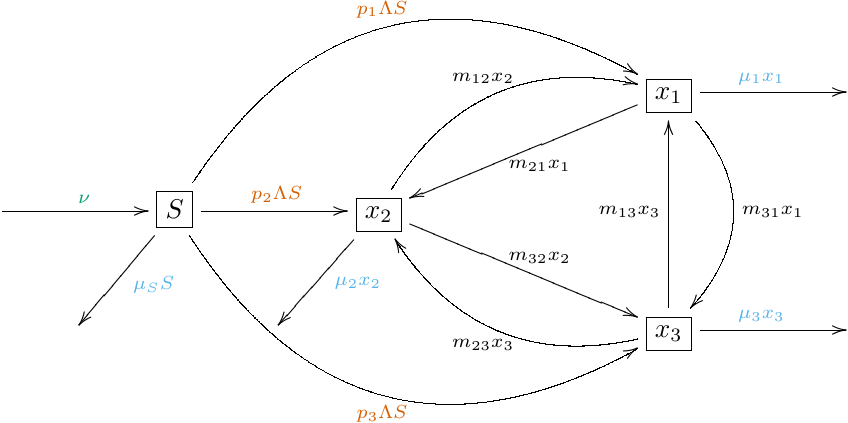}
\end{center}
\caption{Transfer diagram for the most general $Sx_1\cdots x_n$ model
  with $n=3$ non-susceptible compartments ($x_i$).  The subgraph
  excluding $S$ is a complete directed graph on $n$ nodes.  There is
  no flow from any $x_i$ back to $S$. The general class of
  $Sx_1\cdots x_n$ models is defined formally
  by \cref{eq:ODE,eq:inits,eq:FoI,eq:PPP,eq:PITmatrix}.}
\label{fig:transfer.diagram}
\end{figure}

The vast majority of compartmental epidemic models that have been
studied over the last century are generalizations of the simplified KM
model \eqref{eq:KM} and can be represented by a \term{transfer
diagram} like that shown in \cref{fig:transfer.diagram} with a single
susceptible compartment ($S$) and $n$ other compartments
($x_1,\ldots,x_n$) from which there are no flows to $S$.  
\hypertarget{unifsusc}{}
We say these \term{\boldmath$Sx_1\cdots x_n$ models} have the property
of \term{uniform susceptibility} because there is no distinction
(biological, epidemiological or sociological) between susceptibles; in
particular, no current susceptible has ever been infected.
The standard SIR model with vital dynamics corresponds to $n=2$, with
$x_1=I$ and $x_2=R$.  Including a latent period, which requires an
exposed compartment ($E$), yields the \term{SEIR model} with $n=3$,
$x_1=E$, $x_2=I$, and $x_3=R$.  More generally, any number of infected
and immune compartments are possible---for example including
compartments in which individuals are hospitalized, treated, isolated,
or temporarily not infectious---and flows among any or all of these
compartments are possible.  Moreover, the distributions of durations
in any of the compartments can be essentially arbitrarily
distributed \cite{KrylEarn13,HurtKiro19}.

In the thousands of papers written about models in the general class
indicated in \cref{fig:transfer.diagram}---many of which have been used
to support infectious disease management and policy development
\cite{AndeMay91,Heth00,Brau+19}---the typical first steps of analysis are to find
equilibria and determine their local stability or instability.  The
more challenging question of global stability has been resolved much
less frequently.  Indeed, even for the simple SEIR model, GAS of the
EE was not established until 1995 \cite{LiMuldowney1995b}.  Yet only
global analysis can fully characterize the long-term dynamics that can
be displayed by epidemic models.

Substantial advances in global stability analysis have been made in
the last 20 years, including a variety of special cases in the class
of epidemic models depicted in \cref{fig:transfer.diagram}; in
particular, GAS has been proved for a large class of staged
progression models \cite{GuoLi06}, SIR/SEIR models with both serial
and parallel infectious stages \cite{IggidrPlus2007,Koro09},
and a class of models similar to those depicted
in \cref{fig:transfer.diagram} but with the restriction that
all new infections arrive in class $x_1$ \cite{GuoLiShuai2012}.
It has remained unclear, however, how much of the class of $Sx_1\cdots
x_n$ models has the GAS property.  We definitively answer this
question here, and discuss what scope there is to broaden our result
to include more general classes of epidemiological models.

\section*{Results}

We formalize the class of $Sx_1\cdots x_n$ models
(\cref{fig:transfer.diagram}) by writing
\begin{linenomath*}
\begin{subequations}\label{eq:ODE}
\begin{align}
\dSdt &= \susrec - \mu_S S - \FoI S \,, \label{eq:ODE;S}\\
\dxxxdt &= \FoI S \PPP - M \xxx \,, \label{eq:ODE;xvec}
\end{align}
\end{subequations}
\end{linenomath*}
with non-negative initial conditions\footnote{Standard existence and
  uniqueness theory \cite[Chapter I]{Hale1969} ensures that solutions
  of \cref{eq:ODE} exist and are unique for each initial condition.
  In \Methods, we show that solutions approach a bounded set, a
  consequence of which is that solutions exist for all
  $t \ge \tinit$.}
\begin{linenomath*}
\begin{equation}		\label{eq:inits}
S(\tinit) = \Sinit, \qquad
\xxx(\tinit) = \xvecinit \,.
\end{equation}
\end{linenomath*}
In \cref{eq:ODE}, $\susrec$ represents natality, all of which is into
the susceptible compartment.  The \emph{per capita} mortality rate
from class $S$ is $\mu_S$ and from each $x_j$ is $\mu_j$.
The \term{force of infection} is
\begin{linenomath*}
\begin{equation}\label{eq:FoI}
\FoI = \bbbeta^\transpose \xxx,
\hspace{1.0cm} \text{ where } \;\;
\bbbeta =  [ \beta_1 , \cdots , \beta_n ]^{\transpose}
\aand
\xxx = [ x_1 , \cdots , x_n ]^{\transpose}.
\end{equation}
\end{linenomath*}
The transmission rates $\beta_j$ must be non-negative, and to ensure
that transmission is \emph{possible}, it is essential that
$\beta_j > 0$ for at least one $j$, but there are no other
restrictions on the $\beta_j$; thus, transmission from the various
$x_j$ classes can be arbitrarily heterogeneous.  In \cref{eq:ODE;xvec},
\begin{linenomath*}
\begin{equation}\label{eq:PPP}
\PPP =  [ p_1 , \cdots , p_n ]^{\transpose}
\end{equation}
\end{linenomath*}
is the vector of probabilities with which newly infected individuals
move from class $S$ to each 
$x_j$ (so each $p_j \ge 0$ and
$\sum_{j=1}^n p_j = 1$).  The non-susceptible flow rates are specified
by what we call the \term{post-infection transfer (PIT) matrix},
\begin{linenomath*}
\begin{equation}		\label{eq:PITmatrix}
M = \begin{bmatrix} M_{ij} \end{bmatrix}_{n \times n}
\hspace{0.3cm} \text{ with } \;\;
\left\{
\begin{matrix}
	M_{ij} = - m_{ij} \hspace{0.6cm} \text{for $i \ne j$},	\\
	\hspace{-0.6cm} M_{jj} = \mu_j + \sum_{i = 1}^n m_{ij}.
\end{matrix}
\right.
\end{equation}
\end{linenomath*}
Here, the transfer coefficients $m_{ij}$ must be non-negative.
Flows from $x_j$ directly back into $x_j$ have no impact, so we
assume without loss of generality that $m_{jj} = 0$.  The vital
rate parameters $\mu_j$ must be strictly positive.
Consequently, the off-diagonal entries of the PIT matrix are non-positive, the
diagonal entries are strictly positive, and the $j$\textsuperscript{th} column
sum is equal to $\mu_j > 0$.
\hypertarget{M-matrix}{}The column sum positivity and sign pattern
imply that the PIT matrix is a non-singular
\href{https://en.wikipedia.org/wiki/M-matrix}{M-matrix} \cite{BermanPlemmons1984,Fiedler2008},
which necessarily has
a non-negative inverse and eigenvalues that all have
positive real parts.\footnote{An \label{foot:Mmatrix}\term{M-matrix} is a square matrix
  with non-positive off-diagonal elements and eigenvalues with
  non-negative real parts (see \cite[Chapter
    6]{BermanPlemmons1984} or \cite[Chapter 5]{Fiedler2008}).
  If an M-matrix is non-singular, then its inverse is
    non-negative and its eigenvalues all have positive real part; see
    \cite[Theorem 6.2.3, statements (G$_{20}$) and
      (N$_{38}$)]{BermanPlemmons1984}.}

Finally, to prevent the occurrence of formal equilibria that are
biologically meaningless, we assume that no compartments are isolated
from the $S$ compartment, \ie for each $j$, it must be
\emph{possible} for initially susceptible individuals to reach $x_j$
eventually.  To formalize this property, we note that any $n \times n$
matrix $A = \left[ a_{ij} \right]$, with non-negative off-diagonal
entries, induces a directed graph (or \emph{digraph}) based on those
off-diagonal entries, namely the graph with a directed edge from node
$j$ to node $i$ if and only if $a_{ij} > 0$ for $i \ne j$; we consider
the digraph induced in this way by the negative of the PIT matrix,
$-M$.  In the example of \cref{fig:transfer.diagram}, $-M$ induces the
subgraph made up of the edges with black labels.  Thus, we assume that
each node in the transfer diagram (\eg
\cref{fig:transfer.diagram}) can be reached by following a path
beginning at $S$.  Using the terminology of graph theory, $S$
would be called a \term{source node}.  We stress that we do \emph{not}
assume the PIT matrix \eqref{eq:PITmatrix} to be irreducibile, which
would be much more restrictive.

With these assumptions, in \Methods we show that the basic
reproduction number for the general $Sx_1\cdots x_n$
model \eqref{eq:ODE} is
\begin{linenomath*}
\begin{equation}\label{eq:Rzero}
\Rzero = \frac{\susrec}{\mu_S} \bbbeta^{\transpose} M^{-1} \PPP\,,
\end{equation}
\end{linenomath*}
and using the fact that $S$ is a source node, we show that $\Rzero$
is necessarily positive.  \cref{eq:Rzero} is a
generalization\footnote{In addition to being essential for our
asymptotic stability analysis in this paper, the explicit formula for
$\Rzero$ \eqref{eq:Rzero} has practical value on its own, whenever an
expression for $\Rzero$ is needed, since it avoids having to apply the
standard algorithm \cite{vandWatm02} in any specific case.} of the
simple expression $\Rzero=N\beta/\gamma$ for \cref{eq:KM}, since
$\susrec/\mu_S$ is the equilibrium population size, and
$\bbbeta^{\transpose}$ and $M^{-1}
\PPP$ are analogous to $\beta$ and $\gamma^{-1}$.

We find that if $\Rzero\le1$ then there is a unique (disease-free)
equilibrium,
\begin{linenomath*}
\begin{equation}\label{eq:DFE}
\left( \SDFE, \xvecDFE \right) = \left( \frac{\susrec\,}{\mu_S}, \vec{0} \right),
\end{equation}
\end{linenomath*}
whereas if $\Rzero>1$ then there is a second (endemic) equilibrium,
\begin{linenomath*}
\begin{equation}\label{eq:EE}
  \left( \SEE, \xvecEE \right) =
  \left(
    \frac{\susrec/\mu_S}{\Rzero} ,\;
    \susrec \Big( 1 - \frac{1}{\Rzero} \Big) M^{-1} \PPP
    \right)
    \,.
\end{equation}
\end{linenomath*}
The system exhibits a standard transcritical bifurcation
\cite{Stro18}; the two equilibria coincide at $\Rzero=1$ and exchange stabilities
as $\Rzero$ passes through $1$.  We prove, moreover, that if
$\Rzero\le1$ then the DFE \eqref{eq:DFE} is GAS, whereas if $\Rzero>1$
then the EE \eqref{eq:EE} is GAS.

{\bfseries Thus, any $\boldsymbol{Sx_1\cdots x_n}$ model---\emph{any}
epidemic model with a transfer diagram in the class illustrated by
\cref{fig:transfer.diagram}---is globally stable.}  

\section*{Discussion}

Our approach to proving global stability is to construct a Lyapunov
function \cite{LaSalle1976} that works for this very general class of
models \eqref{eq:ODE}.  Compared to previous work, which has depended
on identifying details of specific models, our key innovation is to
recognize that construction of a Lyapunov function for a generic
$Sx_1\cdots x_n$ model can be reduced to a system of linear equations
that can always be solved, together with a set of algebraic
inequalities that we show are always satisfied.

Having settled this question, it is natural to ask whether the class
of models can be expanded further without losing the guarantee of a
GAS equilibrium.

\paragraph{Maternally-acquired immunity.}

Thus far, we have implicitly assumed that individuals are susceptible
when they are born, which is not always true: a period of maternally
acquired immunity is common for many infectious
diseases \cite{AndeMay91,Hethcote2000}.  It follows
from Corollary~4.3 of
\ifpnas
Ref.\,\citen{Thieme1992}
\else
Ref.{\phantom{x}}\raisebox{-5.25pt}{\scalebox{1.45}{\citen{Thieme1992}}}
\fi
that any ``pre-susceptible''
transfer diagram can be pasted onto the left
of \cref{fig:transfer.diagram} without affecting global stability.  In
particular, including maternally acquired immunity still yields a GAS
equilibrium, regardless of how the duration of maternally-acquired
immunity is distributed \cite{KrylEarn13,HurtKiro19}.

\paragraph{Vaccination.}

If a vaccine that provides permanent immunity is given to a proportion
$p$ of newborns, the susceptible recruitment rate is reduced from
$\susrec$ to $\susrec(1-p)$, and the perfectly vaccinated individuals can be
thought of as residing in a vaccinated compartment $V$ (separated
from the potentially infectious classes $x_1,\ldots,x_n$).  Our GAS
result still holds in this situation.  Moreover, GAS follows even if
vaccine-induced immunity decays, \ie if there is a flow from $V$ to
$S$ at rate $\delta V$ for some $\delta>0$ (permanent vaccine-induced
immunity corresponds to $\delta=0$).

Alternatively, or in addition, if susceptibles are vaccinated at a
constant \emph{per capita} rate \ie if there is a flow from $S$ to $V$
at rate $\alpha S$ for some $\alpha>0$, GAS still follows (again, even
if vaccine-induced immunity decays).

Note that for some models involving a ``leaky vaccine''---one that
provides only a temporary \emph{reduction} in susceptibility, rather
than full immunity---it is \emph{possible}  to have \emph{two} stable
equilibria \cite{ArinoMcCluskeyVdD2003}.  However, this bistability is
possible only if there is flow from a vaccinated class to the infected
classes, contradicting our assumption of \hyperlink{unifsusc}{uniform
susceptibility}.

\paragraph{Decay of infection-induced immunity.}

$Sx_1\cdots x_n$ models assume that infection-induced immunity is
permanent, \ie there is no flow back to $S$ after becoming infected.
This assumption of permanent immunity after infection cannot be
dropped.  Some models with decay of immunity do have a GAS equilibrium
\cite{Heth76,Koro04}, however there are simple models with decay of
infection-induced immunity that have an unstable EE accompanied by a
stable periodic orbit \cite{Heth+81}.
\ifpnas
(Ref.\,\citen{Heth+81}
\else
(Ref.{\phantom{x}}\raisebox{-5.25pt}{\scalebox{1.45}{\citen{Heth+81}}}
\fi
ignores
births and deaths, but the result holds with vital dynamics as well
because stable periodic orbits persist under small perturbations;
see Theorem 2 in \S4.1 of
\ifpnas
Ref.\,\citen{Perko2006}.)
\else
Ref.{\phantom{x}}\raisebox{-5.25pt}{\scalebox{1.45}{\citen{Perko2006}}}.)
\fi

\paragraph{Permanent infections.}

The class of $Sx_1\cdots x_n$ models includes diseases from which
individuals never recover to an immune state (\eg tuberculosis, HIV).
Our global stability result still applies in the absense of immunity,
since the only requirement is that individuals never return to the
susceptible class after being infected.

\paragraph{Non-exponential lifetime distributions.}

As drawn, \cref{fig:transfer.diagram} suggests that host lifetimes are
exponentially distributed.  
In fact, since essentially arbitrary stage duration distributions
can be represented using collections of serial and parallel
compartments \cite{KrylEarn13,HurtKiro19}, there is really
no such restriction:
an arbitrary sub-graph can be pasted onto the end of any of the
mortality paths (which point out of the diagram
in \cref{fig:transfer.diagram}) without disrupting our global
stability analysis, as the resulting PIT matrix would still be a
non-singular M-matrix (which can be established using statement
(L$_{32}$) of Theorem 6.2.3 in
\ifpnas
Ref.\,\citen{BermanPlemmons1984}).
\else
Ref.{\phantom{x}}\raisebox{-5.25pt}{\scalebox{1.45}{\citen{BermanPlemmons1984}}}).
\fi

\paragraph{Infection age and class age.}

In their most general model, KM allowed for continuous changes in
infectiousness during an individual's infectious
period \cite{KermMcKe27}, \ie the transmission rate $\beta$ can depend
on how long individuals have been infectious (their \term{infection
age}).  KM expressed this general model as an integro-differential
equation, but there are ODEs in the class of $Sx_1\cdots x_n$ models
that approximate it to any desired degree of accuracy.  The reason is
that any compartment (say, represented by $x_k$) can be subdivided
into a sequence of subcompartments (say,
$x_{k,1},\,\ldots,\,x_{k,K}$), each with a different transmission
rate. All such subcompartmentalized models in which
transmission rate depends on infection-age (or more
generally \term{class-age}) can be represented by \cref{eq:ODE},
implying they each have an equilibrium that is GAS.  In the limit
that the number of subcompartments $K\to\infty$, we obtain an
integro-differential equation like KM's. While the limiting
equation itself is not an ODE, the fact that it can be
approximated arbitrarily closely by GAS
systems formalized by \cref{eq:ODE},
strongly suggests (though does not
prove) that the limiting equation also has a GAS
equilibrium.

\paragraph{Delay.}

Many models (see
\ifpnas
refs.\,\citen{Keyoumu+2023, LiShu-SIAP-2010, LiLiu2014},
\else
refs.{\phantom{x}}\raisebox{-5.25pt}{\scalebox{1.45}{\citen{Keyoumu+2023, LiShu-SIAP-2010, LiLiu2014}}},
\fi
for example) include delays, whereby individuals that leave one group
arrive in another group $\tau > 0$ time units later.  The resulting
delay differential equation (DDE) corresponds to individuals spending
a period of time ($\tau$) in a \emph{hidden} class that is not
explicitly represented in the model, such as a latent class in which
individuals are infected but not infectious.  The delay ($\tau$) could
be the same for all individuals (\ie discrete delay), could take
values from a continuous distribution (\ie distributed delay), or
could be a combination of the two \cite{McCluskey2015SIADS}.

As an example, suppose that upon infection, all individuals experience a
latent period of the same duration ($\tau$), before proceeding to classes
$x_1, \dots, x_n$.  Then \cref{eq:ODE} would become
\begin{linenomath*}
\begin{subequations}\label{eq:delay}
\begin{align}
\dSdt(t) &= \susrec - \mu_S S(t) - \FoI(t) S(t) \,, \label{eq:delay;S}\\
\dxxxdt(t) &= \eee^{- \mulatentsmallL \! \tau} \FoI(t-\tau) S(t-\tau) \PPP - M \xxx(t) \,, \label{eq:delay;xvec}
\end{align}
\end{subequations}
\end{linenomath*}
where $\mulatent$ is the \emph{per capita} mortality rate while in the
latent class.  Thus, a fraction
$\eee^{-\mulatentsmallL \! \tau}$ of those that are infected
at time $t-\tau$ survive to progress to the remaining classes at time
$t$.

In \Methods, we perform global analysis for the ODE
version \eqref{eq:ODE}, using one Lyapunov function for the case
$\Rzero > 1$ and another for the case $\Rzero \le 1$.  Theorem 7.1 of
\ifpnas
Ref.\,\citen{McCluskey2015SIADS}
\else
Ref.{\phantom{x}}\raisebox{-5.25pt}{\scalebox{1.45}{\citen{McCluskey2015SIADS}}}
\fi
allows our result for $\Rzero > 1$ to be extended to include the type
of delay represented in \cref{eq:delay}, and much more general delays
that are represented as mixtures of discrete and continuous
distributions.  A similar approach can be used to extend our result
for $\Rzero \le 1$.

\paragraph{Disease-induced mortality.}

Because the death rate $\mu_j$ in each compartment can be different,
disease-induced mortality is implicitly included in our general
structure.  Indeed, mortality resulting from any characteristic of a
given compartment is covered (\eg post-infection mortality from a
secondary infection, such as fatal pneumonia following influenza
infection).

\paragraph{Proportional scaling.}

In our formulation of the class of $Sx_1\cdots x_n$ models
\eqref{eq:ODE}, we assumed a constant rate of susceptible recruitment
\cite{Earn+00}, rather than a rate that increases in proportion to the
population size $N$.
If we replace $\susrec$ with $\susrec N$ in \cref{eq:ODE} and $\bbbeta$ by
$\bbbeta/N$ in \cref{eq:FoI} then our global stability results are
unaffected.  This follows because a model that has recruitment $\susrec N$
and transmission $\bbbeta/N$---but is otherwise identical to
\cref{eq:ODE}---is formally equivalent to \cref{eq:ODE} after changing
variables to proportions rather than numbers of individuals in each
compartment.  See, \eg \cite[Section 2]{McCluskey2003} for a similar
calculation.

\paragraph{Heterogeneity.}

In the simple KM model \eqref{eq:KM}, the transmission rate $\beta$
represents the product of the contact rate and the probability of infection
upon contact.  In general $Sx_1\cdots x_n$ models \eqref{eq:ODE},
there can be a distinct transmission rate $\beta_j$ associated with
each $x_j$.  Thus, it is straightforward to represent heterogeneity in
the rates at which individuals in each of the $x_j$ classes contact
susceptibles, or in the probabilities of infection upon contact, or both.

\paragraph{Environmental reservoirs.}

The transmission dynamics of some infections (\eg cholera) are driven,
at least in part, by hosts shedding pathogen into the environment,
which becomes an additional source of infection.  Many models that include an
environmental reservoir \cite{TienEarn10,Tien+11} fall into the class
of $Sx_1\cdots x_n$ models \eqref{eq:ODE}.

\paragraph{Nonlinear incidence.}

The models defined by \cref{eq:ODE} assume bilinear incidence, $S\FoI$.
Incidence functions $\inc(S,\FoI)$ that are nonlinear in $S$ or $\FoI$
(or both) are commonly considered, often as a means to approximate
density-dependent mixing
heterogeneities \cite{WilsWorc45,Liu+87,Novo08}.  In this more general
setting, $\FoI$ is still defined by \cref{eq:FoI} but does not
correspond to the force of infection in the usual sense.

Many models with nonlinear incidence have been shown to have a GAS
equilibrium \cite{Koro06,Koro07}, but others exhibit periodic orbits
and bistability \cite{Liu+86,Liu+87}.  Consequently, our results
cannot be generalized to models with \emph{arbitrary} incidence
functions $\inc(S,\FoI)$.
However, our results do generalize to any incidence function that
satisfies some biologically sensible constraints; in particular, to
establish GAS it is sufficient that
(i) $\inc(S,\FoI)$ is zero if either argument is zero,
(ii) $\inc$ is a non-negative increasing function of both $S$ and $\FoI$,
(iii) $\frac{\partial\inc}{\partial \FoI}$ is a non-increasing
function
of $\FoI$ ($\frac{\partial^2 \inc}{\partial \FoI^2} \le 0$) and
(iv) $\frac{\partial\inc}{\partial \FoI}$ is a non-decreasing
function of $S$ ($\frac{\partial^2 \inc}{\partial S \partial \FoI} \ge 0$).
(Note that $\frac{\partial\inc}{\partial \FoI}$ is a measure of the
sensitivity of incidence to changes in the infectious level of the
population.)  The proof of this more general result is substantially
longer than the bilinear case on which we focus here, but the
structure of the proof is similar.

\paragraph{Structured populations.}

The generic $Sx_1\cdots x_n$ model places no restrictions on
stratification of \emph{infected} individuals by age, spatial
location, or other physiological or social factors, but we have
assumed that individuals are indistinguishable when they are
susceptible. It remains to be seen whether the general result we have
presented here can be extended---or to what extent it can be
extended---to include heterogeneous mixing patterns that depend on
stratification of susceptibles into distinct classes.

\paragraph{Non-equilibrium attractors.}

Our focus here has been on the existence of a GAS equilibrium.  Of
course, other types of attractors are possible and can be globally
stable.  Some of the counter-examples we have cited have a stable
periodic orbit, while others exhibit bistability.  Our
approach---based on considering generic digraphs rather than specific
properties of epidemic models---may facilitate additional fruitful
developments that help characterize model structures that necessarily
give rise to one globally stable attractor or several co-existing
attractors, further enhancing our understanding of the ecology and
evolution of infectious diseases.

\typeout{(2) Current \columnsep = \the\columnsep} 

\hypertarget{Methods}{}
\ifpnas
  \relax
\else
  \bigbreak\bigbreak
  \renewcommand{\theequation}{M\arabic{equation}}
  \setcounter{equation}{0}  
  \section*{Methods}\label{sec:Methods}
\fi

\ifpnas
  \matmethods{

\noindent
The intuitive idea that motivates \cref{eq:ODE} is that after
initial infection, individuals may experience a variety of disease
states or situations that can be modelled as passing through
compartments that are connected by a digraph structure.  Such graphs
can always be represented by an M-matrix that we call the PIT matrix
[\cref{eq:PITmatrix}].  This framework includes systems as simple as
the common SEIR model \cite{Korobeinikov2004} and as complex as those
involving multiple infectious groups, quarantine, treatment, quiescence,
and essentially arbitrarily distributed durations in each
compartment.  At one extreme, $M$ could be a diagonal matrix
(though this would require that $\PPP$ be strictly positive), while at
the other extreme all of the off-diagonals of $M$ could be negative.

We restrict attention to models in which there is no return from
post-infection classes to susceptibility because it is
known \cite{Heth+81} that models allowing a return to $S$ \emph{can}
exhibit non-constant periodic solutions, precluding a GAS equilibrium.

We use differential inequalities to establish that solutions that are
initially non-negative remain so (a biological necessity), and also
to show that there is a bounded set that attracts all solutions.

For $\Rzero > 1$, we perform a global stability analysis by
constructing a Lyapunov function based on the Volterra function
$\GG{\theta} = \theta - 1 - \ln (\theta)$, which has been used
extensively for specific subclasses of $Sx_1\cdots x_n$ models.
This leads to a calculation where a set of Lyapunov coefficients must
be found so that certain non-linearities cancel out.  We use the
digraph structure of the post-infection compartments, \ie the M-matrix
structure of the PIT matrix, to show that a suitable choice of
coefficients exists.

\subsection*{Biological well-posedness}\hypertarget{biolwellposed}{}

We write the full state of the system as
$\Xvec=(S,\xxx)$. (Note that we use a double-headed arrow to
distinguish $(n+1)$-dimensional vectors such as $\Xvec$ from
$n$-dimensional vectors such as $\xxx$.)  If $\Xvec$ is non-negative
($\Xvec \in \Rgez^{n+1}$) and one of its components is zero, then the
time-derivative of that variable, as given in
\cref{eq:ODE}, is non-negative.  Consequently, by \cite[Proposition
2.1]{Haddad+2010}, the non-negative orthant ($\Rgez^{n+1}$) is
positively invariant, \ie biologically sensible initial conditions
yield biologically sensible solutions for all time.

\subsection*{Bounds on solutions}	

We denote  the column vector in $\R^n$ with a 1 in each component by
$\ones$, which allows us to write the total population size
conveniently as
\begin{equation}
N(t) = S(t) + \ones^\transpose \xxx(t).
\end{equation}
Then
\begin{equation}
\dNdt \= \dSdt + \ones^\transpose \dxxxdt
	\= \Bigl( \susrec - \mu_S S - \FoI S \Bigr) + \ones^\transpose \Bigl( \FoI S \PPP - M \xxx \Bigr).
\end{equation}
Noting that $\ones^\transpose \PPP = \sum p_j = 1$ and that $\ones^\transpose M$ gives a row
vector containing the column sums of $M$, we have
\begin{equation}
  \dNdt \= \susrec - \mu_S S - \mu_1 x_1 - \cdots
  - \mu_n x_n \;\; \le \;\; \susrec - \mu N,
\end{equation}
where $\mu = \min \left\{ \mu_S, \mu_1, \dots, \mu_n \right\} > 0$.
Consequently, it follows from standard theory for differential inequalities
\cite[Theorem I.6.1]{Hale1969} that
$N(t) \le \frac{\nu}{\mu} + e^{-\mu t}\big(N(\tinit) -
  \frac{\nu}{\mu}\big)$, and hence
\begin{equation}		\label{eq:Bound_N}
\limsup_{t \to \infty} N(t) \le \frac{\susrec}{\mu} \,.
\end{equation}

Since $N$ is asymptotically bounded \eqref{eq:Bound_N}, and
the state variables that sum to $N$ are all non-negative, any linear
combination of the state variables is also asymptotically bounded.
In particular, it follows that there exists $K > 0$ such that
$\FoI(t) \le K$ for sufficiently large $t$, say $t\ge t_K$.
Hence, \cref{eq:ODE;S} implies that for $t\ge t_K$,
\begin{equation}
\dSdt \ge \susrec - \left( \mu_S + K \right) S,
\end{equation}
and therefore all solutions to \cref{eq:ODE} must satisfy
\begin{equation}\label{eq:Sliminf}
\liminf_{t \to \infty} S(t) \ge  \frac{\susrec}{\mu_S + K} > 0.
\end{equation}
Similarly, $\dSdt \le \susrec - \mu_S S$, and so 
\begin{equation}\label{eq:Slimsup}
\limsup_{t \to \infty} S(t) \le \frac{\susrec}{\mu_S}.
\end{equation}

Inequalities \eqref{eq:Bound_N}, \eqref{eq:Sliminf}, and
  \eqref{eq:Slimsup}, together imply that solutions of \cref{eq:ODE}
  asymptotically approach the set
\begin{equation}\label{eq:Dset}
\Dset = \left\{ \left( S, \xxx \right) \in \Rgez^{n+1} : 
  \frac{\susrec}{\mu_S + K} \le S \le \frac{\susrec}{\mu_S}
  \quad\text{and}\quad
  S + x_1 + \cdots + x_n \le \frac{\susrec}{\mu} \right \}.
\end{equation}
More precisely, under the dynamics of
  \cref{eq:ODE} with initial conditions given by
  \eref{eq:inits}, 
  the following statements are true.
\begin{itemize}\itemsep0pt
\item[1]$0 \le S(t) \le N(t) \le \max \left\{ \frac{\susrec}{\mu}, N(\tinit) \right\}$ for all $t \ge \tinit$.
\item[2]The set $\Dset$ is positively invariant and attracts all solutions..
\item[3]The attractor (omega limit set) of each solution is bounded and is contained in $\Dset$.
\item[4]All equilibria are contained in $\Dset$.
\end{itemize}

\subsection*{The disease-free equilibrium (DFE)}\label{sec:DFE}

For all parameter values, there is a unique disease-free equilibrium
\begin{linenomath*}
\begin{equation}\label{eq:XDFE}
\XDFE \= \left( \SDFE, \zero \right) \= \left( \frac{\susrec}{\mu_S}, 0, \dots, 0 \right).
\end{equation}
\end{linenomath*}


\hypertarget{subsection:Rzero}{}
\newcommand{\Rzerosecname}{The basic reproduction number $\Rzero$}
\subsection*{\Rzerosecname}\label{sec:Rzero}

We use the next generation matrix method as detailed in
\ifpnas
Ref.\,\citen{VdDWatmough2002}
\else
Ref.{\phantom{x}}\raisebox{-5.25pt}{\scalebox{1.45}{\citen{VdDWatmough2002}}}
\fi
to calculate the basic reproduction number.  Let
\begin{linenomath*}
\begin{equation}
\mathcal F = \FoI S \PPP
\aaand
\mathcal V = M \xxx.
\end{equation}
\end{linenomath*}
Then 
\begin{linenomath*}
\begin{equation}\label{eq:FandV}
F = \frac{\partial \mathcal F}{\partial \xxx} \left( \XDFE \right)
	= \frac{\susrec}{\mu_S} \PPP \bbbeta^{\transpose}
\aaand
V = \frac{\partial \mathcal V}{\partial \xxx} \left( \XDFE \right) = M.
\end{equation}
\end{linenomath*}
The next generation matrix $\mathsf{N}$ is
\begin{linenomath*}
\begin{equation}	\label{NextGen}
\mathsf{N} \= F V^{-1} \= \frac{\susrec}{\mu_S} \PPP \bbbeta^{\transpose} M^{-1} \,,
\end{equation}
\end{linenomath*}
and $\Rzero$ is the spectral radius of this matrix
\cite{Diek+90,VdDWatmough2002}.  Noting that $\bbbeta^\transpose M^{-1}$
is a row vector, it follows from \cref{NextGen} that each column of
$\mathsf{N}$ is a scalar multiple of $\PPP$, and so $\mathsf{N}$ has rank
one.  Consequently, at most one eigenvalue of the next generation matrix
is non-zero.  \cref{NextGen} also implies that 
${\mathsf{N}} \PPP
= \frac{\susrec}{\mu_S}\PPP \big(\bbbeta^{\transpose} M^{-1} \PPP\big)
= \frac{\susrec}{\mu_S} \big( \bbbeta^{\transpose} M^{-1} \PPP\big)\PPP$,
and therefore $\PPP$ is an eigenvector of $\mathsf{N}$ corresponding
to the eigenvalue $\lambdaNaught =
\frac{\susrec}{\mu_S} \bbbeta^{\transpose} M^{-1} \PPP$.
\ifpnas
In \hyperlink{lambdaNaught.positive}{\bfseries\slshape Proof that the eigenvalue $\lambdaNaught$ of $\mathsf{N}$ is positive} we show
that $\lambdaNaught>0$,
\else
We show in \supp that $\lambdaNaught$ is positive,
\fi
from which it follows that it is the spectral radius of $\mathsf{N}$,
and hence $\Rzero = \lambdaNaught$, as stated in \cref{eq:Rzero}.


\hypertarget{subsection:EE}{}
\newcommand{\EEsecname}{The endemic equilibrium (EE)}
\subsection*{\EEsecname}\label{sec:EE}

We now look for an equilibrium $\XEE = \left( \SEE, \xvecEE \right)$
with $\SEE, \FoIEE > 0$, where $\FoIEE = \bbbeta^{\transpose} \xvecEE$.
We will see that such an equilibrium exists (and is unique) if and only if
$\Rzero>1$.

If such an $\XEE$ exists, then
$\dSdt ( \XEE )=0$, and so it follows from
Equations~\eqref{eq:ODE;S} and \eqref{eq:XDFE}
that
\begin{equation}\label{eq:SEE<SDFE}
\SEE \= \frac{\susrec}{\mu_S + \FoIEE} \;<\; \SDFE.
\end{equation}
In addition, \cref{eq:ODE;xvec} becomes
$\dxxxdt ( \XEE ) = \zero$, which implies
$\xvecEE = \FoIEE \SEE M^{-1} \PPP$.  Left multiplying by
$\bbbeta^{\transpose}$ yields
\begin{linenomath*}
\begin{equation}		\label{eq:FoIStar}
\FoIEE = \FoIEE \SEE \bbbeta^{\transpose} M^{-1} \PPP.
\end{equation}
\end{linenomath*}
Cancelling $\FoIEE$ and recalling the definition of $\Rzero$ [\cref{eq:Rzero}],
we find
\begin{linenomath*}
\begin{equation}\label{eq:SEE}
\SEE = \frac{\SDFE}{\Rzero}.
\end{equation}
\end{linenomath*}
Since $\SEE < \SDFE$ \eqref{eq:SEE<SDFE}, 
\cref{eq:SEE} implies that $\Rzero > 1$, precluding the possibility of
sub-threshold endemic equilibria
\ifpnas
(see Ref.\,\citen{VdDWatmough2002}).
\else
(see Ref.{\phantom{x}}\raisebox{-5.25pt}{\scalebox{1.45}{\citen{VdDWatmough2002}}}).
\fi
Using $\dSdt(\XEE)=0$ again, we find $\FoIEE \SEE = \susrec - \mu_S \SEE$,
allowing the expression for $\xvecEE$ to be rewritten as
\begin{linenomath*}
\begin{equation}\label{eq:xvecEE}
\xvecEE \= \left( \susrec - \mu_S \SEE \right) M^{-1} \PPP \= \susrec \left( 1 - \frac{1}{\Rzero} \right) M^{-1} \PPP.
\end{equation}
\end{linenomath*}
Since $M^{-1}$ and $\PPP$ are non-negative, it follows that $\xvecEE$
is also a non-negative vector; moreover, since \cref{eq:xvecEE}
implies that the non-zero vector $\PPP$ is a scalar multiple of
$M \xvecEE$, it follows further that $\xvecEE$ is not the zero vector.
In fact, our assumption that each node in the transfer diagram can be
reached by a path beginning at $S$ implies that \emph{none} of the
components of $\xvecEE$ can be zero, so $\xvecEE \in \Rgtz^n$;
\ifpnas
see
\hyperlink{xvecEE.positive}{\bfseries\slshape Proof that $\xvecEE$ is strictly positive}.
\else
see \supp.
\fi
Summarizing, we have:
\begin{lemma}
  For any $\Rzero>0$, \cref{eq:ODE} has a unique disease-free
  equilibrium,
  $\XDFE = (\SDFE,\xvecDFE) = \left( {\susrec}/{\mu_S},\, \zero
  \right)$.
  If $\Rzero \le 1$, then $\XDFE$ is the only non-negative equilibrium
  for \cref{eq:ODE}.  If $\Rzero > 1$, then there is also a (unique)
  endemic equilibrium
  $\XEE = \left( \SEE, \xvecEE \right) \in \Rgtz^{n+1}$.
\end{lemma}

\subsection*{The Volterra function $\G$}\label{sec:g}

In our definition and analysis of Lyapunov functions below, we make
judicious use of the Volterra function\footnote{We adopt this name in honour of
Vito Volterra's work \cite{Volterra1928, Volterra1939} on predator-prey
models, in which he calculated first integrals that can be written in terms of $\G$.}
\begin{linenomath*}
\begin{equation}\label{eq:g}
\GG{\theta} = \theta - 1 - \ln \theta,
\end{equation}
\end{linenomath*}
and several of its properties, which we list here:

\begin{description}
\hypertarget{g1=0}{}
\item[P1]$\GG{\theta}$ has a unique minimum at
  $\theta=1$, with  $\GG{1} = 0$.
\hypertarget{g_lemma_2}{}
\item[P2]\label{eq:g_lemma_2}
  If $A_1, A_2, B_1, B_2 > 0$ then
\begin{linenomath*}
\begin{equation}
\left( A_1 - A_2 \right) \left( B_1 - B_2 \right)
	= \GG{A_1 B_1} - \GG{A_1 B_2} - \GG{A_2 B_1} + \GG{A_2 B_2}.
\end{equation}
\end{linenomath*}

\hypertarget{g_lemma_new}{}
\item[P3]\label{eq:g_lemma_new}Suppose $A, B > 0$, 
$T_i\ne0$ for $i = 1, \dots, i_{\max}$,
$\sgn(T_i) =  \sgn(T_i^*)$, and $\sum_i T_i^* = 0$.
If $t_i = \frac{T_i}{T_i^*}$ for each $i$ then
\begin{linenomath*}
\begin{equation}
\left( A - B \right) \sum_i T_i
	= \sum_i T_i^* \left[ \GG{A t_i} - \GG{B t_i} \right].
\end{equation}
\end{linenomath*}

\end{description}
A proof of \hyperlink{g1=0}{Property \textbf{P1}} is elementary.
\hyperlink{g_lemma_2}{Property \textbf{P2}} appears as Lemma~2.3 in
\ifpnas
Ref.\,\citen{McCluskey2015SIADS}.
\else
Ref.{\phantom{x}}\raisebox{-5.25pt}{\scalebox{1.45}{\citen{McCluskey2015SIADS}}}.
\fi
A proof of \hyperlink{g_lemma_new}{Property \textbf{P3}} is given
\ifpnas
in \hyperlink{P3.proof}{\bfseries\slshape Proof that the Volterra function $\G$ satisfies \upshape Property \textbf{P3}}.
\else
in \supp.
\fi

\subsection*{Lyapunov analysis for $\Rzero \le 1$}\label{sec:R0_small}

We define
\begin{linenomath*}
\begin{equation}
\LiapDFE_S = \Rzero \SDFE \GG{\frac{S}{\displaystyle\SDFE}}
\aand
\LiapDFE_{\xxx} = \SDFE \bbbeta^{\transpose} M^{-1} \xxx,
\end{equation}
\end{linenomath*}
and let
\begin{linenomath*}
\begin{equation}	\label{eq:UUU}
\LiapDFE = \LiapDFE_S + \LiapDFE_{\xxx}.
\end{equation}
\end{linenomath*}
As noted after \cref{eq:PITmatrix},
$M^{-1}$ is non-negative, and therefore $\LiapDFE_{\xxx} \ge 0$.  The
function $\LiapDFE$ is defined for $S > 0$, $x_1, \dots, x_n \ge 0$.
We will show that $\LiapDFE$ is a Lyapunov function for \cref{eq:ODE}
  if $\Rzero\le1$, and hence that all solutions in $\Rgez^{n+1}$ tend
to $\XDFE$.

We begin by differentiating $\LiapDFE_S$ with respect to time as points
$\left( S, \xxx \right)$ move along solutions to \cref{eq:ODE},
obtaining
\begin{linenomath*}
\begin{equation}
\frac{\dee{}\LiapDFE_S}{\dee{t}} 
	\= \frac{\partial \LiapDFE_S}{\partial S \;} \dSdt
	\= \Rzero \left( 1 - \frac{\SDFE}{S} \right) \left[ \susrec - \mu_S S - \FoI S \right].
\end{equation}
\end{linenomath*}
Inserting $\susrec = \mu_S \SDFE$, we write
\begin{linenomath*}
\begin{equation}
\frac{\dee{}\LiapDFE_S}{\dee{t}} 
= - \frac{\mu_S \Rzero}{S} \Bigl( S - \SDFE \Bigr)^2 
- \Rzero \FoI S + \Rzero \FoI \SDFE.
\end{equation}
\end{linenomath*}
Differentiating $\LiapDFE_{\xxx}$ with respect to time gives
\begin{linenomath*}
\begin{equation}
\begin{aligned}
\frac{\dee{}\LiapDFE_{\xxx}}{\dee{t}} 
	&= \SDFE \bbbeta^{\transpose} M^{-1} \dxxxdt				
			\\
	&= \SDFE \bbbeta^{\transpose} M^{-1} \PPP \, \FoI S
				\quad - \quad \SDFE \bbbeta^{\transpose} M^{-1} M \xxx
			\\
	&= \Rzero \FoI S \quad - \quad \SDFE \FoI.
\end{aligned}
\end{equation}
\end{linenomath*}
Adding $\displaystyle\frac{\dee{}\LiapDFE_S}{\dee{t}}$ and
$\displaystyle\frac{\dee{}\LiapDFE_{\xxx}}{\dee{t}}$ together to get $\displaystyle\dUdt$, 
and then recalling $\Rzero\le 1$,
we obtain 
\begin{linenomath*}
\begin{equation}
\begin{aligned}
\dUdt \;=\; - \frac{\mu_S \Rzero}{S} \Bigl( S - \SDFE \Bigr)^2 + \left( \Rzero - 1 \right) \FoI \SDFE
	\;\le\; 0,
\end{aligned}
\end{equation}
\end{linenomath*}
where necessary conditions for equality include $S = \SDFE$.

Let $\Bset$ be the set on which \scalebox{0.8}{$\dUdt$} is zero and
let $\Aset$ be the largest invariant subset of $\Bset \cap \Dset$.
Note that $\Aset$ is bounded since $\Dset$ as defined
in \cref{eq:Dset} is bounded.  Since $S = \SDFE$ throughout $\Bset$
(and hence $\Aset$), and $\Aset$ is invariant, it follows that
$\dSdt \equiv 0$ in $\Aset$.  Thus, in $\Aset$,
\begin{linenomath*}
\begin{equation}
0 \= \dSdt \= \susrec - \mu_S \SDFE - \FoI \SDFE \= - \FoI \SDFE.
\end{equation}
\end{linenomath*}
The only possibility is that $\FoI \equiv 0$.  Consequently, within
$\Aset$, \cref{eq:ODE;xvec} becomes
\begin{linenomath*}
\begin{equation}		\label{eq:dxxxdtU}
\dxxxdt = - M \xxx.
\end{equation}
\end{linenomath*}
Since all of the eigenvalues of $M$ have positive real part, the only
solution $\xxx(t)$ of \cref{eq:dxxxdtU} that is bounded for all $t \in \R$
(and hence for which the corresponding $\Xvec(t) = \left( \SDFE, \xxx(t) \right)$
lies in $\Aset$ for all $t \in \R$) is the zero solution, $\xxx(t) \equiv \zero$.
Therefore,
\begin{linenomath*}
\begin{equation}
\Aset
\= \left\{ (\SDFE,\zero) \right\}
\= \left\{ \XDFE \right\}.
\end{equation}
\end{linenomath*}
Thus, by LaSalles' Invariance Principle \cite{LaSalle1976}, any
solution to \cref{eq:ODE} starting in the domain of $\displaystyle\LiapDFE$ will tend to
$\XDFE$.  Since any solution that starts with $S=0$ immediately moves
to a state with $S > 0$, we see that any solution starting in
$\Rgez^{n+1}$ tends to $\XDFE$, giving:
\begin{theorem}
  If $\Rzero \le 1$, then \cref{eq:ODE} has a globally asymptotically stable
  disease-free equilibrium, $\XDFE$.
\end{theorem}


\hypertarget{subsection:R0_big}{}
\newcommand{\LyapunovBigsecname}{Lyapunov analysis for $\Rzero > 1$}
\subsection*{\LyapunovBigsecname}\label{sec:R0_big}

We define
\begin{linenomath*}
\begin{equation}
\LiapEE_S = \SEE \,\GG{\frac{S}{\SEE}}
\aand
\LiapEE_k = \xEEk \,\GG{\frac{x_k}{\xEEk}}
\;\textrm{ for } k = 1, \dots, n,
\end{equation}
\end{linenomath*}
and let
\begin{linenomath*}
\begin{equation}	\label{eq:VVV}
\LiapEE = \LiapEE_S + \sum_{k=1}^n a_k \LiapEE_k,
\end{equation}
\end{linenomath*}
where the \term{Lyapunov coefficients}
$a_1, \dots, a_n \ge 0$, so $\LiapEE : \Rgtz^{n+1} \to \Rgez$.
Specific values of the $a_k$ will be determined below.  If each $a_k$
is positive, then $\LiapEE$ achieves its minimum value of zero only at
the endemic equilibrium $\XEE$.  However, if some of the $a_k$ are
zero, then there is a larger set on which $\LiapEE$ is zero;
additional calculations that are required for this situation are
presented
\ifpnas
in \hyperlink{some.ak.zero}{\bfseries\slshape Analysis details if some Lyapunov coefficients are zero when $\Rzero>1$}.
\else
in \supp.
\fi

To establish that $\LiapEE$ is a Lyapunov function (thereby showing
that $\XEE$ is GAS), we must show that \scalebox{0.8}{$\dVdt$} is
non-positive.  To that end, it is convenient to define
\begin{linenomath*}
\begin{equation}	\label{eq:s_y_k}
s = \frac{S}{\SEE}
\quad\; \textrm{ \rm and } \quad\;
y_k = \frac{x_k}{\xEEk} \quad
\text{for } k = 1, \dots, n.
\end{equation}
\end{linenomath*}
Differentiating $\LiapEE_S$ with respect to time gives
\begin{linenomath*}
\begin{equation}
\frac{\dee{\LiapEE_S}}{\dee{t}}
	\= \frac{\partial \LiapEE}{\partial S} \dSdt
	\= \left( 1 - \frac{\SEE}{S} \right) \left[ \susrec - \mu_S S - \sum_{j=1}^n \beta_j x_j S \right].
\end{equation}
\end{linenomath*}
Using the above definitions of $s$ and $y_k$,
the fact that $\G(1)=0$
  [\hyperlink{g1=0}{Property \textbf{P1}}], and
  \hyperlink{g_lemma_new}{Property \textbf{P3}} with $A=1$,
  $B=\frac{\SEE}{\displaystyle S}$, 
  $\{T_i\}$
  given by
  $\left\{ \susrec, -\mu_S S, -\beta_1 x_1 S, \dots, -\beta_n x_n S
  \right\}$, and $T_i^*$ given by the equilibrium value of the
  corresponding $T_i$, we obtain
\begin{linenomath*}
\begin{equation}\label{eq:dVSdt}
\frac{\dee{\LiapEE_S}}{\dee{t}}
	= - \susrec \,\GG{\frac{1}{s}} - \mu_S \SEE \,\GG{s} + \sum_{j=1}^n \beta_j \xEEj \SEE \left[ \GG{y_j} - \GG{y_j s} \right].
\end{equation}
\end{linenomath*}

Next, for $k = 1, \dots, n$, we differentiate $\LiapEE_k$ with respect to
time, obtaining
\begin{linenomath*}
\begin{equation}
\begin{aligned}
\frac{\dee{\LiapEE_k}}{\dee{t}}
	\= \frac{\partial \LiapEE_k}{\partial x_k} \frac{\dee{x_k}}{\dee{t}}
	\= \left( 1 - \frac{\xEEk}{x_k} \right)
		\left[ p_k \sum_{j=1}^n \beta_j x_j S + \sum_{j=1}^n m_{kj} x_j - \left( \mu_k + \sum_{l=1}^n m_{lk} \right) x_k \right].
\end{aligned}
\end{equation}
\end{linenomath*}
Noting that $\frac{\dee{x_k}}{\dee{t}}$ is zero at $\XEE$, leads to
$\left( \mu_k + \sum_{l=1}^n m_{lk} \right) \xEEk = p_k \sum_{j=1}^n \beta_j \xEEj \SEE + \summkj$, and hence
\begin{linenomath*}
\begin{equation}
\begin{aligned}
\frac{\dee{\LiapEE_k}}{\dee{t}}
	&= \left( 1 - \frac{\xEEk}{x_k} \right)
		\left[ p_k \sum_{j=1}^n \beta_j x_j S + \sum_{j=1}^n m_{kj} x_j
		- \left( p_k \sum_{j=1}^n \beta_j \xEEj \SEE + \summkj \right) \frac{x_k}{\xEEk} \right]	\\
	&= \left( 1 - \frac{1}{y_k} \right)
		\left[ p_k \sumjbeta \left( y_j s - y_k \right)
			+ \summkj \left( y_j - y_k \right) \right]					\\
	&= p_k \sumjbeta \left( 1 - \frac{1}{y_k} \right)\left( y_j s - y_k \right)
			+ \summkj \left( 1 - \frac{1}{y_k} \right) \left( y_j - y_k \right).
\end{aligned}
\end{equation}
\end{linenomath*}
Again using the fact that $\G(1)=0$, but now exploiting
\hyperlink{g_lemma_2}{Property \textbf{P2}}, we have
\begin{linenomath*}
\begin{equation}\label{eq:dVkdt}
\frac{\dee{\LiapEE_k}}{\dee{t}}
	= p_k \sumjbeta \left[ \GG{y_j s} - \GG{y_k} - \GG{\frac{y_j s}{y_k}} \right]
		+ \summkj \left[ \GG{y_j} - \GG{y_k} - \GG{\frac{y_j}{y_k}} \right].
\end{equation}
\end{linenomath*}

From the definition of $\LiapEE$ in \cref{eq:VVV}, we have
\begin{equation}\label{eq:dVdt}
\dVdt = \frac{\dee{\LiapEE_S}}{\dee{t}} + \sum_{k=1}^n a_k \frac{\dee{\LiapEE_k}}{\dee{t}}.
\end{equation}
We now use \cref{eq:dVSdt,eq:dVkdt} to expand the right-hand side
of \cref{eq:dVdt}.  We then collect all terms in which the argument of
$\G$ is simply $y_j$ or $y_k$, and rewrite those terms such that the
separate sums over $j$ and $k$ are combined to form a single sum over
$\ell$; thus, all terms involving $\GG{y_j}$ or $\GG{y_k}$ are replaced
with terms involving $\GG{y_\ell}$, yielding
\begin{linenomath*}
\begin{equation}\label{eq:dVdt_1}
\begin{aligned}
\dVdt &= - \susrec \,\GG{\frac{1}{s}} - \mu_S \SEE \,\GG{s}
			- \sum_{k=1}^n \sum_{j=1}^n \left[ a_k p_k \beta_j \xEEj \SEE \,\GG{\frac{y_j s}{y_k}}
					+ a_k m_{kj} \xEEj \,\GG{\frac{y_j}{y_k}} \right]							\\
	&\hspace{2.0cm} + \sumjbeta \,\GG{y_j s} \left[ - 1 + \sum_{k=1}^n a_k p_k \right]				\\
	&\hspace{2.0cm} + \sum_{\ell=1}^n \GG{y_\ell} \left[ \beta_\ell \xEEl \SEE
			- a_\ell p_\ell \sumjbeta + \sum_{k=1}^n a_k m_{k\ell} \xEEl - \sum_{j=1}^n a_\ell m_{\ell j} \xEEj \right].
\end{aligned}
\end{equation}
\end{linenomath*}

We now work to show that the coefficients $a_k \ge 0$ can be chosen so
that the final two lines of \cref{eq:dVdt_1} vanish.  Let
\begin{equation}\label{eq:Bl}
A_{kj} = m_{kj} \xEEj \aaand
B_\ell = \beta_\ell \xEEl \SEE,
\end{equation}
and note that $\sum_{\ell=1}^n B_\ell = \FoIEE \SEE \ne 0$.  
Suppose now that it is possible to choose
the $a_k$ such that, for each $\ell$, the coefficient of $\GG{y_\ell}$
on the final line of \cref{eq:dVdt_1} (appearing in square brackets)
vanishes, \ie such that
\begin{linenomath*}
\begin{equation}\label{eq:Coef_A}
0 = B_\ell - a_\ell p_\ell \FoIEE \SEE + \sum_{k=1}^n a_k A_{k\ell} - a_\ell
\sum_{j=1}^n A_{\ell j},
\qquad \ell \in \left\{ 1, \dots, n \right\}.
\end{equation}
\end{linenomath*}
Summing over $\ell$, the terms involving the $A$'s cancel, giving
\begin{linenomath*}
\begin{equation}
\begin{aligned}
0 \= \sum_{\ell=1}^n B_\ell -  \sum_{\ell=1}^n a_\ell p_\ell \FoIEE \SEE 
    \ifpnas\relax\else\\\fi
	\= \left( 1 - \sum_{\ell=1}^n a_\ell p_\ell \right) \FoIEE \SEE.
\end{aligned}
\end{equation}
\end{linenomath*}
Since $\FoIEE \SEE$ is non-zero, we see that any choice of the $a_k$
that solves \cref{eq:Coef_A} for each $\ell$, will give
$\left( 1 - \sum_{\ell=1}^n a_\ell p_\ell \right) = 0$,
thereby eliminating the second line of \cref{eq:dVdt_1}.
Thus, eliminating the third line of \cref{eq:dVdt_1}
automatically eliminates the second as well\footnote{This cancellation is
  consistent with a phenomenon that arises in the global analysis of
  many models, where there are fewer Lyapunov coefficients to choose
  than there are restrictions to satisfy, and yet a valid choice
  exists; see \cite{McCluskey2006, McCluskey2008a} for example.},
leaving only the first line, which is non-positive (as desired)
since $g \ge 0$ (\hyperlink{g1=0}{Property \textbf{P1}}).  It
remains to show that the $a_k$ can be chosen such that
\cref{eq:Coef_A} is satisfied.

We write
\begin{linenomath*}
\begin{equation}
\vec a = [ a_1 , a_2 , \cdots,  a_n ]^{\transpose}
\aand
\vec B = [ B_1 , B_2 , \cdots , B_n ]^{\transpose},
\end{equation}
\end{linenomath*}
so that \cref{eq:Coef_A} can be expressed in matrix form as
\begin{linenomath*}
\begin{equation}		\label{eq:QaB}
Q \vec a = \vec B,
\end{equation}
\end{linenomath*}
where
\begin{linenomath*}
\begin{equation}
Q = \begin{bmatrix}
p_1 \FoIEE \SEE + \sum_{j \ne 1} A_{1j} & -A_{21} & \cdots & -A_{n1}		\\
-A_{12} & p_2 \FoIEE \SEE + \sum_{j \ne 2} A_{2j} & \cdots & -A_{n2}		\\
\vdots & \vdots & \ddots & \vdots	\\
-A_{1n} & -A_{2n} & \cdots & p_n \FoIEE \SEE + \sum_{j \ne n} A_{nj}
\end{bmatrix}.
\end{equation}
\end{linenomath*}
The column sums of $Q$ can be written as a column vector,
\begin{equation}
Q^{\transpose} \ones = \FoIEE \SEE \PPP,
\end{equation}
the components of which are non-negative and not all zero (since
$\FoIEE>0$, $\SEE>0$, $\PPP\ge0$, and $\sum_k p_k=1$).
Combining
this fact with the sign pattern of $Q$ and our assumption that each
node in the transfer diagram can be reached by a path beginning at
$S$, it follows from statement (L$_{32}$) of \cite[Theorem
6.2.3]{BermanPlemmons1984} that $Q^{\transpose}$, and hence $Q$, is a
non-singular M-matrix.
Consequently, $Q$ satisfies all fifty enumerated statements of
\cite[Theorem 6.2.3]{BermanPlemmons1984}, including statement
(N$_{38}$), which implies that each of the entries of $Q^{-1}$ is
non-negative.  Rewriting \cref{eq:QaB} as $\vec a = Q^{-1} \vec B$,
and recalling that the components of $\vec B$ as defined in
\cref{eq:Bl} are non-negative, it follows that $\vec a$ is
non-negative.  Moreover, since at least one $B_l$ is non-zero (as
$\FoIEE\ne0$) \cref{eq:QaB} implies that $\vec a$ cannot be the
zero vector.

Since \cref{eq:QaB} is equivalent to \cref{eq:Coef_A},
  the choice $\vec{a}=Q^{-1}\vec{B}$ successfully eliminates
the second and third lines of \cref{eq:dVdt_1}, and therefore
\begin{linenomath*}
\begin{equation}
\begin{aligned}\label{eq:dVdtnonpos}
\dVdt \ =\  - \susrec \,\GG{\frac{1}{s}} - \mu_S \SEE \,\GG{s}
			- \sum_{k=1}^n \sum_{j=1}^n \left[ a_k p_k \beta_j \xEEj \SEE \,\GG{\frac{y_j s}{y_k}}
					+ a_k m_{kj} \xEEj \,\GG{\frac{y_j}{y_k}} \right]	
\ \le\  0.
\end{aligned}
\end{equation}
\end{linenomath*}
Here, non-positivity follows from the fact that the Volterra
  function ($\G$), the state variables ($S,x_k$), and the parameters
  ($\susrec,\mu_S,a_k,p_k,\beta_k,m_{kj}$), are all non-negative.  In
  addition, given that $\GG{\theta}=0$ if and only if $\theta=1$
  [\hyperlink{g1=0}{Property P1}], and $\susrec>0$, a necessary
  condition for equality in \cref{eq:dVdtnonpos} is that $s=1$ or,
  equivalently, $S = \SEE$.

Similar to the argument in the previous section, we complete the proof
that $\XEE$ is GAS using LaSalles' Invariance Principle
\cite{LaSalle1976}.  Let $\Bset \subset \Rgtz^{n+1}$ be the set on
which \scalebox{0.8}{$\dVdt$} is zero.  Let $\Aset$ be the largest
invariant subset of $\Bset \cap \Dset$, and note that
$\Aset\subset\Dset$ is bounded.  Then in the set $\Aset$, we have
$S=\SEE$.  Hence, by the invariance of $\Aset$, we also have
$\dSdt \equiv 0$.  Thus,
\begin{linenomath*}
\begin{equation}
0 \= \dSdt \= \susrec - \mu_S \SEE - \FoI \SEE.
\end{equation}
\end{linenomath*}
The only solution is $\FoI \equiv \FoIEE$.
Thus, the equation for $\dxxxdt$ becomes
\begin{linenomath*}
\begin{equation}		\label{eq:dxxxdt}
\dxxxdt = \FoIEE \SEE \PPP - M \xxx.
\end{equation}
\end{linenomath*}
Since all of the eigenvalues of $M$ have positive real part, the only
solution of \cref{eq:dxxxdt} that is bounded for all $t \in \R$ is the
constant solution $\xxx \equiv \FoIEE \SEE M^{-1} \PPP = \xvecEE$.  It
follows that $\xxx(t) \equiv \xvecEE$ for any solution
$(\SEE,\xxx(t))$ in $\Aset$, and so
\begin{linenomath*}
\begin{equation}
\Aset
\= \left\{ (\SEE,\xvecEE) \right\}
\= \big\{ \XEE \big\}.
\end{equation}
\end{linenomath*}
By LaSalles' Invariance Principle, any solution to \cref{eq:ODE} 
with strictly positive initial conditions
tends to $\XEE$, giving: 
\begin{theorem}
  If $\Rzero > 1$, then \cref{eq:ODE} has a globally asymptotically
  stable endemic equilibrium, $\XEE$.
\end{theorem}

\subsection*{Further technical details}

\hypertarget{xvecEE.positive}{}
\subsubsection*{Proof that $\xvecEE$ is strictly positive}

When determining the equilibria (see \hyperlink{subsection:EE}{{\bfseries\EEsecname}}),
we showed that $\xvecEE\in \Rgez^n$ is a non-zero vector and, therefore, at least
one component of $\xvecEE$ is positive.  Suppose some of the
components of $\xvecEE$ are zero.  Without loss of generality, we may
assume that
\begin{linenomath*}
\begin{equation}	\label{eq:xvecEE_blocks}
\xvecEE = \begin{bmatrix} \vec q_{\LL} \\ \zero_{n-\LL} \end{bmatrix},
\end{equation}
\end{linenomath*}
where $\vec q_{\LL} \in \Rgtz^{\LL}$ for some
$\LL \in \left\{ 1, \dots, n-1 \right\}$ and $\zero_{n-\LL}$ is the zero
vector in $\R^{n-\LL}$.
The PIT matrix can be written as
\begin{linenomath*}
\begin{equation}	\label{eq:M_blocks}
M = \begin{bmatrix} B_{11} & B_{12} \\ B_{21} & B_{22} \end{bmatrix},
\end{equation}
\end{linenomath*}
where $B_{11}$ is $\LL \times \LL$, $B_{22}$ is $(n-\LL) \times (n-\LL)$
and the blocks $B_{12}$ and $B_{21}$ are non-positive.
Additionally, $\PPP$ can be written as
\begin{linenomath*}
\begin{equation}	\label{eq:P_blocks}
P = \begin{bmatrix} \PPP_1 \\ \PPP_2 \end{bmatrix},
\end{equation}
\end{linenomath*}
where $\PPP_1 \in \Rgez^{\LL}$ and $\PPP_2 \in \Rgez^{n-\LL}$.  Then,
at the endemic equilibrium, \cref{eq:ODE;xvec} can be written as
\begin{linenomath*}
\begin{equation}	\label{eq:M_blocks1}
\begin{bmatrix} B_{11} & B_{12} \\ B_{21} & B_{22} \end{bmatrix}
   \begin{bmatrix} \vec q_{\LL} \\ \zero_{n-\LL} \end{bmatrix}
=
\FoIEE \SEE \begin{bmatrix} \PPP_1 \\ \PPP_2 \end{bmatrix},
\end{equation}
\end{linenomath*}
\goodbreak
\noindent and therefore
\begin{linenomath*}
\begin{equation}	\label{eq:M_blocks2}
B_{21} \vec q_{\LL} = \FoIEE \SEE \PPP_2.
\end{equation}
\end{linenomath*}
Noting that $\vec q_{\LL}$ is a strictly positive vector and that
$\FoIEE \SEE > 0$, while $B_{21}$ is non-positive and $\PPP_2$ is
non-negative, it follows that the only way \cref{eq:M_blocks2} can
be satisfied is for $B_{21}$ and $\PPP_2$ to both be zero.

Since $\PPP_2 $ is zero, the transfer diagram does not include flow
from $S$ directly to any of the nodes $x_{\LL+1}, \dots, x_n$.  Since
$B_{21}$ is zero, there is no flow from nodes $x_1, \dots, x_{\LL}$
into nodes $x_{\LL+1}, \dots, x_n$.  This contradicts our assumption
that each node in the transfer diagram can be reached by a directed
path beginning at $S$.  Thus, it must be the case that $\xvecEE$ is
strictly positive.\qed

\hypertarget{P3.proof}{}
\subsubsection*{Proof that the Volterra function $\G$ satisfies \hyperlink{g_lemma_new}{\upshape Property \textbf{P3}}}

\begin{linenomath*}
\begin{subequations}
\begin{align}
\sum_i T_i^* \left[ \GG{A t_i} - \GG{B t_i} \right]
&= \sum_i T_i^* \left[ \Bigl( A t_i - 1 - \ln A - \ln t_i \Bigr) - \Bigl( B t_i - 1 - \ln B - \ln t_i \Bigr) \right]	\\
&= \sum_i T_i^* \left[ \Bigl( A t_i - \ln A \Bigr) - \Bigl( B t_i - \ln B \Bigr) \right]	\\
&= \left( A - B \right) \sum_i T_i^* t_i + \left( \ln B - \ln A \right) \sum_i T_i^* 	\\
&= \left( A - B \right) \sum_i T_i, \label{P3proof;4}
\end{align}
\end{subequations}
\end{linenomath*}
as desired, where the final step to conclude \eqref{P3proof;4} follows
from the definition of $t_i$ and the fact that the $T_i^*$ sum to zero.
\qed

\hypertarget{some.ak.zero}{}
\subsubsection*{Analysis details if some Lyapunov coefficients are zero when $\Rzero>1$}

In \hyperlink{subsection:R0_big}{{\bfseries\LyapunovBigsecname}}, we
introduced the Lyapunov coefficients $a_k\ge0$, but assumed for convenience
that $a_k>0$ for all $k$ in \cref{eq:VVV}.  We now show---via reduction to
a representative submodel---that it is sufficient for one of the $a_k$
to be positive.

Suppose some of the $a_k$ are zero.  For instance, suppose $a_1, \dots, a_{\rr} > 0$
and $a_{\rr+1}, \dots, a_n = 0$ with $1 \le \rr < n$.
(If $\rr = n$, then all of the Lyapunov coefficients are positive,
which is the ``main'' case.)

Write $Q$, $\vec a$ and $\vec B$ in block form as
\begin{linenomath*}
\begin{equation}
Q = \begin{bmatrix} Q_{11} & Q_{12} \\ Q_{21} & Q_{22} \end{bmatrix},
\qquad
\begin{bmatrix} \vec a_{\rr} \\ \zero_{n-\rr} \end{bmatrix}
\aand
\begin{bmatrix} \vec B_1 \\ \vec B_2 \end{bmatrix},
\end{equation}
\end{linenomath*}
where $Q_{11}$ is $\rr \times \rr$, $Q_{22}$ is $(n-\rr) \times (n-\rr)$,
$\vec a_{\rr} \in \Rgtz^{\rr}$, $\vec B_1 \in \Rgez^{\rr}$,
$\vec B_2 \in \Rgez^{n-\rr}$, and $\zero_{n-\rr}$ is the zero vector in
$\R^{n-\rr}$.  Note that the entries of $Q_{12}$ and
$Q_{21}$ are all non-positive.  \cref{eq:QaB} can be written as
\begin{linenomath*}
  \begin{equation}
Q_{11} \vec a_{\rr} = \vec B_1
\qquad\text{and}\qquad
\label{eq:QaB2}
Q_{21} \vec a_{\rr} = \vec B_2.
\end{equation}
\end{linenomath*}
Based on the signs of the entries in $Q_{21}$, $\vec a_{\rr}$ and $\vec B_2$, \cref{eq:QaB2} implies $Q_{21}$ is
a zero matrix and $\vec B_2$ is a zero vector.

The fact that $Q_{21}$ and $\vec B_2$ are zero is consequential when
we consider the original differential equation \cref{eq:ODE}.  The
$(i,j)$ off-diagonal entry of $Q$ is $- A_{ji} = - m_{ji} \xEEi$.
Since $\xvecEE \in \Rgtz^n$, the fact that $Q_{21} = 0$ implies the
matrix $M$ has a block of zeros in the transpose position
(corresponding to the location of $Q_{12}$).  The fact that $\vec B_2
= \zero_{n-\rr}$ implies $\FoI = \sum_{j=1}^{\rr} \beta_j x_j$.

Combining these facts, it follows that the differential equations for
$\dSdt, \frac{\dee{} x_1}{\dee{t}\;}, \dots, \frac{\dee{} x_{\rr}}{\dee{t}\;}$ do not depend on $x_{\rr+1}, \dots, x_n$.
Thus, it is understandable that the Lyapunov function would be a function of $S$, $x_1$, \dots, $x_{\rr}$, and
not depend on $x_{\rr+1}, \dots, x_n$, as is the case if $a_{\rr+1} = \dots = a_n = 0$.  In such a case, the Lyapunov
function could be used to show that the subsystem of $\left( S, x_1, \dots, x_{\rr} \right)$ tends to an equilibrium.
Then, the subsystem of $\left( x_{\rr+1}, \dots, x_n \right)$ can be studied as a non-autonomous linear system,
driven by $\FoI S \PPP$, and would also tend to an equilibrium.

Thus, no generality is actually lost by our assumption in the main text that each $a_k>0$.

\hypertarget{lambdaNaught.positive}{}
\subsubsection*{Proof that the eigenvalue $\lambdaNaught$ of $\mathsf{N}$ is positive}\label{M_inv_positive}

When calculating $\Rzero$ (see
  \hyperlink{subsection:Rzero}{{\bfseries\Rzerosecname}}) we showed
that the next generation matrix $\mathsf{N}$ [\cref{NextGen}] has at
most one non-zero eigenvalue.  We show here
that the eigenvalue $\lambdaNaught$ (equal to $\frac{\susrec}{\mu_S}
\bbbeta^{\transpose} M^{-1} \PPP$) is, in fact, positive.
Consequently, as stated \hyperlink{subsection:Rzero}{earlier}, this
positive eigenvalue is the spectral radius of $\mathsf{N}$,
\ie $\lambdaNaught$ is equal to $\Rzero$ \cite{Diek+90,VdDWatmough2002}.

Recalling that the vital rates $\susrec$ and $\mu_S$ are positive, and
that $\bbbeta$, $\PPP$ and $M^{-1}$ are non-negative [see the
  paragraph containing \cref{eq:FoI,eq:PPP,eq:PITmatrix}], it follows
that $\lambdaNaught$ is also non-negative.  It
  remains to show that $\lambdaNaught$ is non-zero.

Observe that
\begin{equation}
\int_0^\infty \eee^{-M \tau} \dtau
\= -M^{-1} \eee^{-M \tau} \Big|_{\tau = 0}^\infty
\= -M^{-1} \left( \lim_{\tau \to \infty} \eee^{-M \tau} \right) + M^{-1} \eee^{-M \tau} \Big|_{\tau = 0}.
\end{equation}
Since the real parts of the eigenvalues of $M$ are all positive
[see \cref{foot:Mmatrix}], the limit above gives the zero matrix.  For the
evaluation at $\tau=0$, we note that the exponential of the zero
matrix is the identity, and so
\begin{equation}	\label{M_inv_int}
M^{-1} = \int_0^\infty \eee^{-M \tau} \dtau.
\end{equation}
Now, since the off-diagonal entries of $M$ are non-positive (as
discussed after \cref{eq:PITmatrix}), we can
choose $\rsrs > 0$ such that $H = \rsrs I - M$ is non-negative.
Then $H^\ell$ is non-negative for any integer $\ell \ge 0$.  Note
that $H$ has the same off-diagonal entries as $-M$, and so
the digraph induced by $H$ is the same as the one induced by
$-M$.

In the digraph induced by $H$ (or by $-M$), a directed edge from 
node $j$ to node $i$ would imply $m_{i j} > 0$.  This can also
be called a path of length one.  A path of length $\ell=2$
from node $j$ through node $\ill$ to node $i$ would imply $m_{i \ill},
m_{\ill\!j} > 0$.  The $(i,j)$ entry of $H^2$ is the sum $\sum_k m_{i
k} m_{k j}$, which, due to the non-negativity of $H$ would be at least
as large as the product $m_{i \ill} m_{\ill\!j} > 0$.  Thus, we see
that a path of length $\ell=2$ implies the corresponding entry of $H^2$
is positive.

More generally, if the digraph induced by
$H$ contains a path of length $\ell \ge 2$ from node $j$ to node
$i$, then $H$ contains $\ell$ positive off-diagonal entries
$m_{{i_{k}}i_{k-1}}$ (for $k = 1,\dots,\ell$) with $i_0 = j$ and $i_\ell =
i$.  The $(i,j)$ entry of $H^\ell$ is a sum of non-negative terms that
includes the positive product $m_{i_\ell i_{\ell-1}} \cdots\, m_{{i_1}{i_0}}$.
Thus, the existence of a path of length $\ell \ge 1$ implies the $(i,j)$-entry
of $H^\ell$ is positive.

If an index $j$ is in the support both of $\PPP$ and $\bbbeta$, then
we say there is a path of length $\ell=0$ from node $j$ to node $i$
(with $i=j$).  The matrix $H^0$ is the identity matrix, and so we can
now assert that the existence
of a path of length $\ell \ge 0$ implies that the
$(i,j)$-entry of $H^\ell$ is positive.

Using the definition of the matrix exponential as a power
series \cite{Coppel1965}
(\ie $\eee^{H \tau} = \sum_{\ell=0}^\infty \frac{1}{\ell!} H^\ell \tau^\ell$),
it follows that the existence of a directed path of any length from node
$j$ to node $i$ implies that the $(i,j)$-entry of $\eee^{H \tau}$ is also
positive for all $\tau > 0$.

Choose integers $j$ and $i$ such that $p_j, \beta_i > 0$ and such that
there exists a path from $x_j$ to $x_i$.  (Our assumption that each node
in the original transfer diagram can be reached by a path starting at $S$
implies that such a choice exists.)  Let $h(\tau) > 0$ be the $(i,j)$-entry of
$\eee^{H \tau}$.  Then
\begin{equation}		\label{beta_H_p_pos}
\bbbeta^\transpose \eee^{H \tau} \PPP \;\; \ge \;\; \beta_i h(\tau) p_j \;\; > \;\; 0.
\end{equation}

Since $-M = -\rsrs I+H$, we have
$\eee^{-M \tau} = \eee^{-\rsrs \tau I + H \tau}
= \eee^{-\rsrs \tau} \eee^{H \tau}$.  Combining this with \cref{M_inv_int},
we obtain
\begin{equation}
\begin{aligned}
\bbbeta^{\transpose} M^{-1} \PPP
	\= \int_0^{\infty} \eee^{-\rsrs \tau} \, \bbbeta^\transpose \eee^{H \tau} \PPP \, \dtau
	\;\; \ge \;\; \int_0^{\infty} \eee^{-\rsrs \tau} \, \beta_i h(\tau) p_j \, \dtau
	\;\; > \;\; 0,
\end{aligned}
\end{equation}
from which it follows that $\lambdaNaught$ is positive, and therefore
$\Rzero = \lambdaNaught$,
proving the validity of \cref{eq:Rzero}.
\qed

  }
  \showmatmethods{} 
  \acknow{Both authors were supported by Discovery Grants from the Natural Sciences and
Engineering Research Council of Canada (NSERC).}
  \showacknow{} 
\else
\fi

  \bibliography{EarnMcCluskey2025_final}
  \ifpnas
  \else
    \bibliographystyle{vancouver}
  \fi

\end{document}